\pdfoutput=1

\documentclass[prd,amsmath,superscriptaddress,%
               nobibnotes,nofootinbib,preprintnumbers,
               twocolumn]{revtex4-1}

\usepackage{placeins}

\usepackage{graphicx}
\usepackage[utf8]{inputenc}
\usepackage[T1]{fontenc}
\usepackage{lmodern} 

\usepackage{ifthen}

\usepackage{braket}

\usepackage{multirow}

\usepackage{url}

\usepackage{verbatim}

\graphicspath{{Figures/}}

\usepackage[normalem]{ulem}
\usepackage[usenames]{color}
\definecolor{DarkRed}{rgb}{0.5,0.0,0.0}
\definecolor{DarkGreen}{rgb}{0.0,0.5,0.0}
\definecolor{DarkBlue}{rgb}{0.0,0.0,0.5}
\definecolor{Magenta}{rgb}{1.0,0.0,1.0}
\definecolor{DarkMagenta}{rgb}{0.5,0.0,0.5}
\definecolor{Orange}{rgb}{1.0,0.5,0.0}
\definecolor{DarkOrange}{rgb}{0.8,0.3,0.0}
\definecolor{DarkCyan}{cmyk}{1.0,0.0,0.0,0.5}
\definecolor{Brown}{cmyk}{0.0,0.8,1,0.6}

\usepackage{soul}

\usepackage{listings}

\definecolor{lstbkgdcolor}{gray}{0.85}

\lstdefinestyle{Pstyle}{language=sh,
        xleftmargin=1.5\parindent,xrightmargin=1.5\parindent,
        columns=fixed,basicstyle=\ttfamily,basewidth=0.5em,
        frame=single,
        backgroundcolor=\color{lstbkgdcolor},
        gobble=1
        }

\lstdefinestyle{Fstyle}{language=[95]Fortran,
        xleftmargin=1.5\parindent,xrightmargin=1.5\parindent,
        columns=fixed,basicstyle=\ttfamily,basewidth=0.5em,
        frame=single,
        backgroundcolor=\color{lstbkgdcolor},
        gobble=1
        }

\lstdefinestyle{Cstyle}{language=C++,
        xleftmargin=1.5\parindent,xrightmargin=1.5\parindent,
        columns=fixed,basicstyle=\ttfamily,basewidth=0.5em,
        frame=single,
        backgroundcolor=\color{lstbkgdcolor},
        gobble=1
        }

\lstnewenvironment{snippet}{\pagebreak[2]}{}
\lstnewenvironment{Psnippet}{\pagebreak[2]\lstset{style=Pstyle}}{\pagebreak[2]}
\lstnewenvironment{Fsnippet}{\pagebreak[2]\lstset{style=Fstyle}}{\pagebreak[2]}
\lstnewenvironment{Csnippet}{\pagebreak[2]\lstset{style=Cstyle}}{\pagebreak[2]}

\newcommand{\refapp}[2][sec:]{the Appendix}
\newcommand{\Refapp}[2][sec:]{The Appendix}

\newcommand{\ifmulticol}[2]{%
  \ifthenelse{\lengthtest{1.9\columnwidth<\textwidth}}{#1}{#2}%
}

\newcommand{\insertfig}[2][]{%
  \hspace*{\stretch{1}}
  \ifthenelse{\NOT\equal{#1}{}}%
    {\includegraphics[keepaspectratio,width=#1\columnwidth]{#2}}%
    {\ifthenelse{\lengthtest{1.9\columnwidth<\textwidth}}%
      {\includegraphics[keepaspectratio,width=1.05\columnwidth]{#2}}%
      {\includegraphics[keepaspectratio,width=0.70\columnwidth]{#2}}%
    }%
  \hspace*{\stretch{1}}
}

\newcommand{\insertwidefig}[2][]{%
  \hspace*{\stretch{1}}
  \ifthenelse{\NOT\equal{#1}{}}%
    {\includegraphics[keepaspectratio,width=#1\textwidth]{#2}}%
    {\ifthenelse{\lengthtest{1.9\columnwidth<\textwidth}}%
      {\includegraphics[keepaspectratio,width=0.80\textwidth]{#2}}%
      {\includegraphics[keepaspectratio,width=0.99\textwidth]{#2}}%
    }%
  \hspace*{\stretch{1}}
}

\newcommand{\insertdoublefig}[3][0.49]{%
    \includegraphics[keepaspectratio,width=#1\textwidth,
                     height=0.35\textheight]{#2}
    \hspace{\stretch{1}}
    \includegraphics[keepaspectratio,width=#1\textwidth,
                     height=0.35\textheight]{#3}
}

\newcommand{\p}{\mathbf{p}}

\newcommand{\micrOMEGAs}{\texttt{micrOMEGAs}}  

\usepackage{hyperref}

\begin{document}

\title{Gamma rays from muons from WIMPs: Implementation of radiative muon decays for dark matter analyses}


\author{Andre Scaffidi}
\email[]{andre.scaffidi@adelaide.edu.au}
\affiliation{
  ARC Center of Excellence for Particle Physics at the Terascale \& CSSM,
  Department of Physics,
  University of Adelaide, 
  Adelaide SA 5005,
  Australia}
  
  \author{Katherine Freese}
\email[]{ktfreese@umich.edu}
\affiliation{Department of Physics, University of Michigan, Ann Arbor
, MI 48109, USA }
\affiliation{
  Oskar Klein Centre for Cosmoparticle Physics, Stockholm University,
  SE-106 91 Stockholm, Sweden}
\affiliation{Nordita,
  KTH Royal Institute of Technology and Stockholm University,
  SE-106 91 Stockholm, Sweden}  

\author{Jinmian Li}
\email[]{jinmian.li@adelaide.edu.au}
\affiliation{
  ARC Center of Excellence for Particle Physics at the Terascale \& CSSM,
  Department of Physics,
  University of Adelaide, 
  Adelaide SA 5005,
  Australia}

\author{Christopher Savage}
\email[]{chris@savage.name}
\affiliation{Nordita,
  KTH Royal Institute of Technology and Stockholm University,
  SE-106 91 Stockholm, Sweden}  
  
\author{Martin White}
\email[]{martin.white@adelaide.edu.au}
\affiliation{
  ARC Center of Excellence for Particle Physics at the Terascale \& CSSM,
  Department of Physics,
  University of Adelaide, 
  Adelaide SA 5005,
  Australia}

\author{Anthony G. Williams}
\email[]{anthony.williams@adelaide.edu.au}
\affiliation{
  ARC Center of Excellence for Particle Physics at the Terascale \& CSSM,
  Department of Physics,
  University of Adelaide, 
  Adelaide SA 5005,
  Australia}

\date{\today}
 


\preprint{ADP-16-13/T968}
\preprint{Nordita-2016-24}

\begin{abstract}
Dark matter searches in gamma ray final states often make use of the fact that photons can be produced from final state muons. Modern Monte Carlo generators and DM codes include the effects of final state radiation from muons produced in the dark matter annihilation process itself, but neglect the $\mathcal{O}$(1\%) radiative correction that arises from the subsequent muon decay. After implementing this correction we demonstrate the effect that it can have on dark matter phenomenology by considering the case of dark matter annihilation to four muons via scalar mediator production. We first show that the AMS-02 positron excess can no longer easily be made consistent with this final state once the Fermi-LAT dwarf limits are calculated with the inclusion of radiative muon decays, and we next show that the  Fermi-LAT galactic centre gamma excess can be improved with this final state after inclusion of the same effect. We provide code and tables for the implementation of this effect in the popular dark matter code \texttt{micrOMEGAs}, providing a solution for any model producing final state muons.


\end{abstract} 

\maketitle



\noindent

\section{Introduction}
\label{sec:Intro}

The failure of the Standard Model (SM) of particle physics to adequately explain dark matter has prompted the development of a large number of particle candidates beyond the SM. Weakly Interacting Massive Particles (WIMPs) are excellent candidates.  Astrophysical and collider searches have heavily constrained these models through either non-observations, or through the interpretation of tentative anomalies as WIMP signals.

An important class of observation sensitive to the particle physics of the WIMP is the search for a gamma ray flux reaching Earth from a dark matter-dominated region of the universe such as a distant dwarf galaxy, or the Galactic Centre. One may obtain photons in WIMP
annihilation in a variety of ways, including direct production (via loop-mediated processes), virtual internal bremstrallung, or the decay of SM annihilation products (e.g. gauge bosons, pions, heavy leptons). The case of dark matter annihilating primarily to leptons (``leptophilic dark matter'' \cite{PhysRevD.79.083528,Bi2009168}) is particularly interesting, since one would not expect to see such WIMPs in hadron collider or direct search experiments. 

In this paper, we look further at the case of gamma rays produced via dark matter annihilation into muonic final states. Modern Monte-Carlo event generators and DM codes include the effects of final state photon radiation (FSR) from muons produced in WIMP annihilation processes, but ignore the radiative decay of the muon in which a photon is emitted from the decaying muon, the intermediate W boson or the final state electron (see Fig. \ref{fig::RadDecay}). To make this clear, final state radiation refers specifically to photons emitted off a muon in the final state of the annihilation process, \textit{not} the muon decay itself, which occurs as a separate process once the muon has propagated. 
\footnote{The muon in the final state of the DM annihilation process is treated a a stable asymptotic state. This is of course an approximation, because the muon eventually decays (with a lifetime of $2.2 \times 10^{-6}$ s enhanced by appropriate time dilation) and so it is really a resonance with a finite width. }  

In this paper, we show that radiative muon decay can play a significant role in dark matter phenomenology. By revisiting the theoretical results for this process, we compute revised gamma ray spectra using the \tt PYTHIA 8.175 \rm Monte Carlo (MC) generator \cite{sjostrand2008jhep05}. These results are generally applicable for any dark matter model that allows WIMP annihilation to muons, and we provide tabulated spectra of results for use with standard dark matter codes such as \micrOMEGAs \cite{Belanger:2001fz}.

In particular we apply our results to the cosmic ray electron-positron anomaly identified by the AMS-02~\cite{Accardo:2014lma}, PAMELA~\cite{Adriani:2008zr,Adriani:2013uda} and Fermi-LAT experiments~\cite{Abdo:2009zk} (with earlier indications coming from HEAT~\cite{Barwick:1997ig,DuVernois:2001bb,Beatty:2004cy}). Assuming that the large AMS-02 signal arises purely from a large annihilation cross-section (rather than for example, an additional boost factor due to an overdense WIMP environment), a previous study ~\cite{Lopez:2015uma} determined that almost all pure SM final states can be excluded as good AMS-02 candidates at greater than $2\sigma$ C.L. by the \texttt{Pass 7} Fermi-LAT dwarf constraints on gamma ray emission~\cite{Ackermann:2013yva}. The one notable exception was the case of a 4-$\mu$ final state produced via the pair production and decay of an unknown mediator particle $\phi$, for which the analysis could not be completed due to the absence of the radiative correction to the muon decay:
\begin{align*}
\chi\chi\rightarrow \phi\phi \rightarrow \mu^+ \mu^- \mu^+ \mu^- \;.
\end{align*} 
We thus revisit this case in this paper using the newest Fermi likelihoods from the \texttt{Pass 8} data \cite{Ackermann:2015zua}.   This work makes use of the standard cosmic ray propagation model known as MED \cite{Boudaud:2014dta}; this is the model that best fits the B/C ratio in the cosmic ray data.

Finally, it is worthwhile to revisit the case of the Fermi-LAT galactic centre data, which is hypothesised to show an excess consistent with WIMP annihilation~\cite{Vitale:2009hr,Hooper:2011ti,Goodenough:2009gk,Gordon:2013vta,Abazajian:2014fta,Daylan:2014rsa} (although several astrophysical explanations have been put forward~\cite{PhysRevLett.116.051103,Bartels:2015aea,Cholis:2015dea,Boyarsky:2010dr,Petrovic:2014uda,Carlson:2014cwa}). 
Previous studies have indicated that the prompt gamma ray spectrum from leptonic final states fails to accurately reproduce the excess in the energy spectrum~\cite{Gordon:2013vta}, since the shape is hard to reconcile with the distribution arising from leptons (which turns out to be peaked towards the kinematic endpoint at the dark matter mass). Given the softer spectrum arising from the radiative decay contribution, we investigate if the prompt gamma ray distribution provides a better fit to the data once the extra effect is included. Note that we limit our study to the distribution of gamma rays produced from the annihilation process itself; a further softening of the distribution can be expected from the effects of charged lepton propagation through the Galactic medium which will broaden the range of masses consistent with the excess relative to those we obtain~\cite{Lacroix:2014eea}. 

Our paper is structured as follows. In Section~\ref{sec:background} we briefly review the necessary theoretical background concerning muon decay. We compute and present the revised gamma ray spectra in Section~\ref{sec:spectra} before investigating the AMS-02 and Fermi-LAT results in Sections~\ref{sec:ams} and~\ref{sec:fermi}. We present conclusions in Section~\ref{sec:conclusions}, and provide information in Appendix~\ref{sec:micro} for users of \micrOMEGAs.
\begin{figure}
		\includegraphics[scale=0.3]{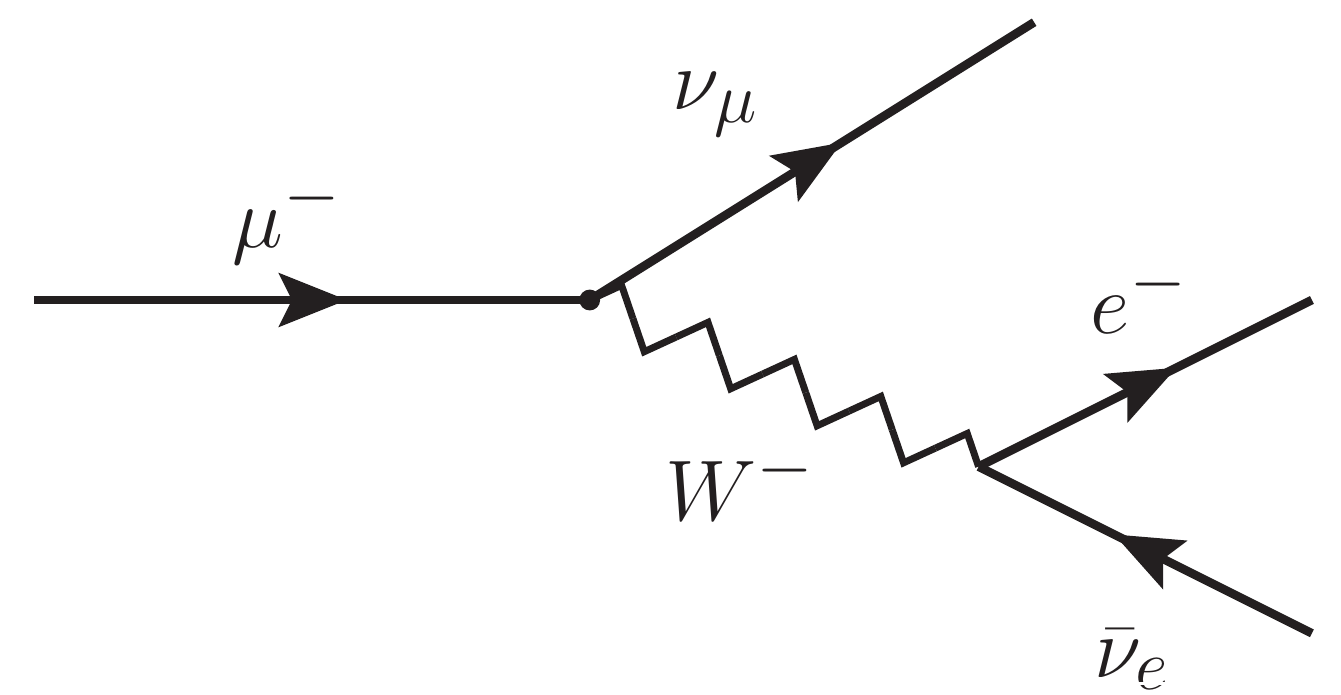}
		\caption{Michel muon decay proceeds most of the time. Shown is 			the particle content of the decay; charges of the final state 			leptons and $W$ boson depend on whether initial state is $\mu^-$ 			or $\mu^+$.}
		\label{fig::standardDecay}
\end{figure}

~
\section{Background: Radiative muon decay}
\label{sec:background}

The standard (Michel) decay of the muon is shown in Fig. \ref{fig::standardDecay}.

\begin{figure}[ht]
		\includegraphics[width=0.23\textwidth]{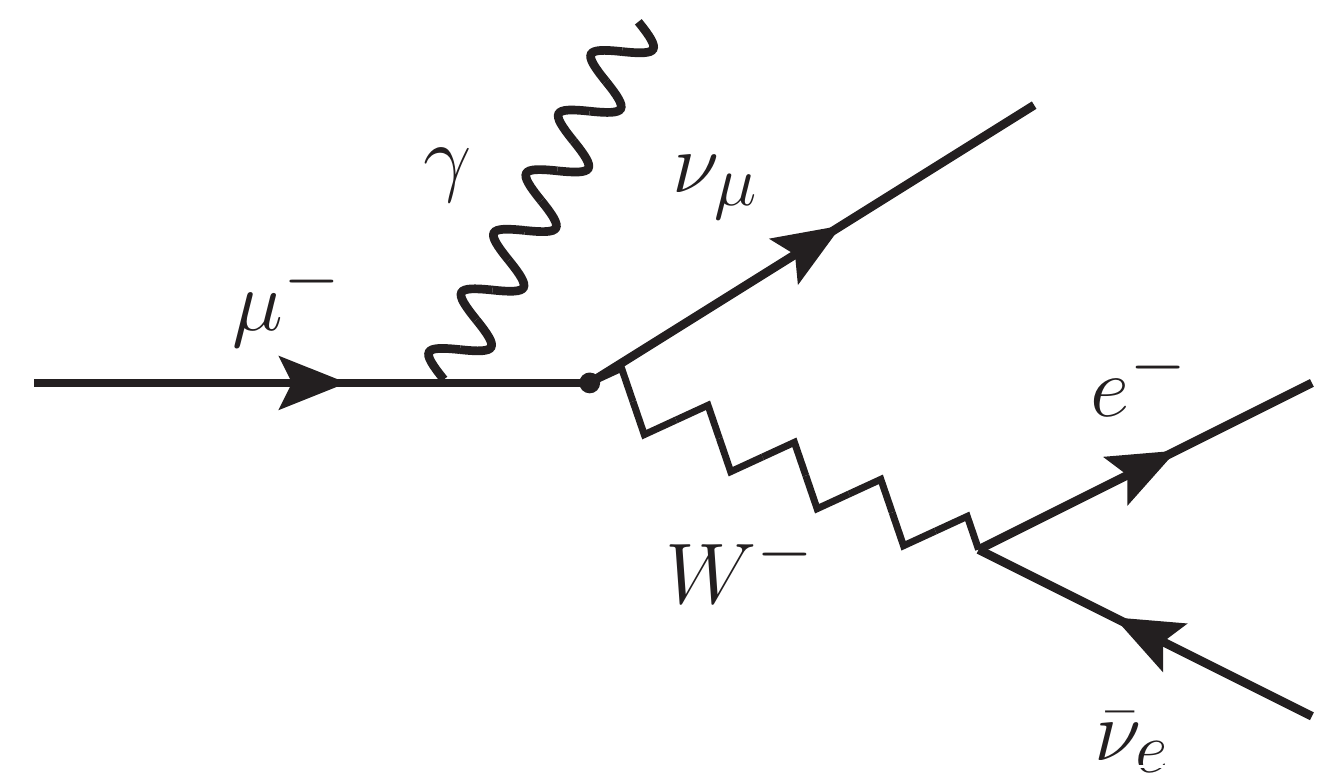}
		\label{fig::RadDecay1}
		~
		\includegraphics[width=0.23\textwidth]{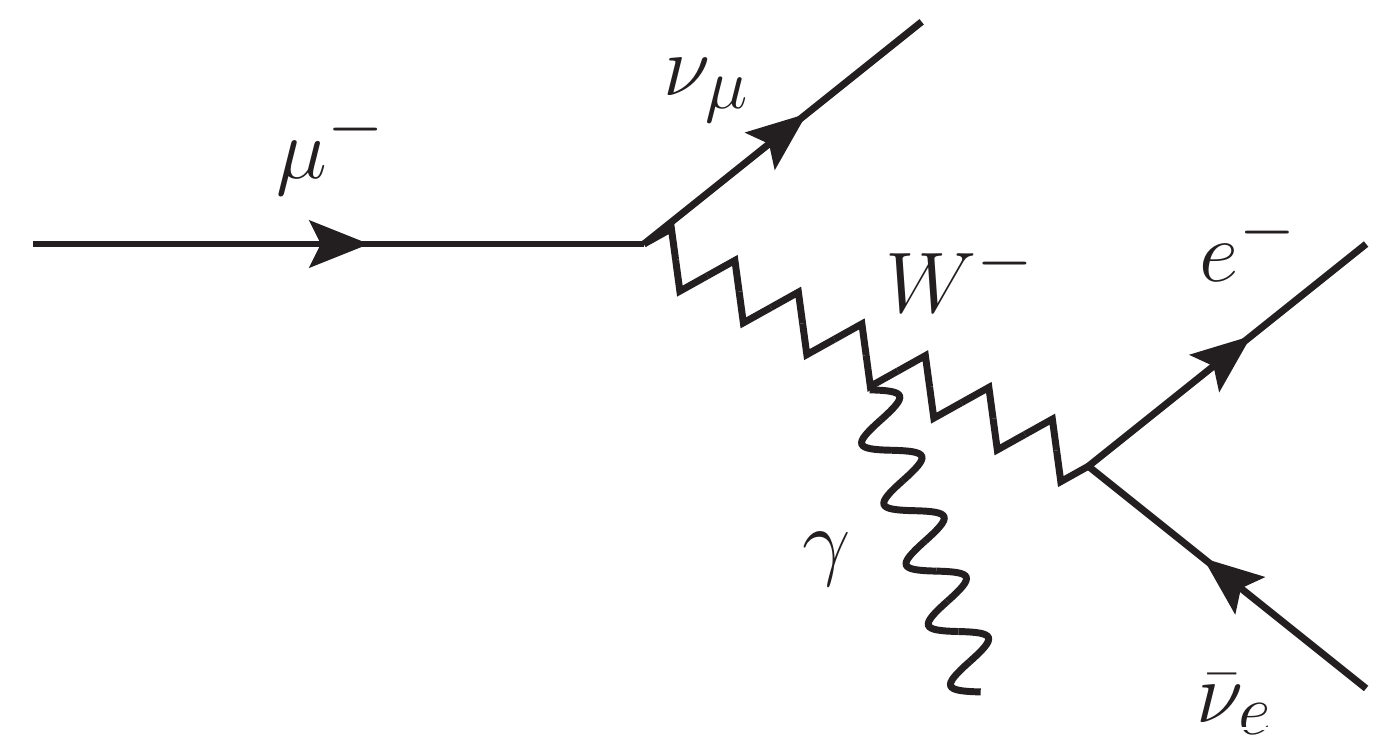}
		\label{fig::RadDecay3}
		~
		\includegraphics[width=0.23\textwidth]{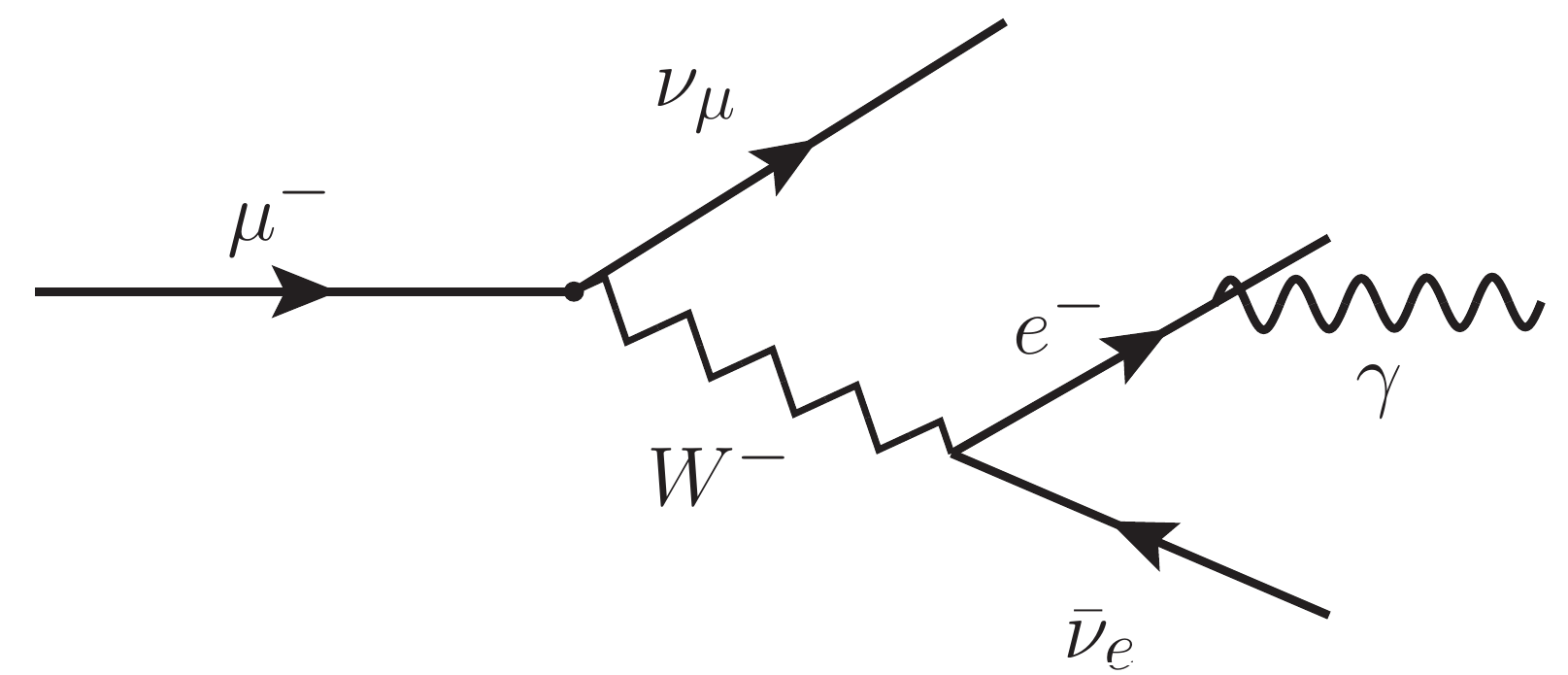}
		\label{fig::RadDecay2}
	\caption{Radiative contributions to the muon decay. An on shell muon can decay, in the process emitting a photon from itself, the intermediate W boson or the electron. }
	\label{fig::RadDecay}
\end{figure}

At the next order in $\alpha_\text{EM}$ one obtains a radiative correction to the muon decay shown in Fig.~\ref{fig::RadDecay} in which a photon can be emitted off either the muon, electron or intermediate vector boson. 
The experimentally measured branching ratio (BR) is dependent on the lower photon energy threshold of the detector. At present, the radiative BR has only been measured for a lower threshold of $E_\gamma>10$ MeV. The best measurement is $\text{BR}_{\mu\rightarrow e\nu\bar{\nu}\gamma} = 1.4\pm 0.4\%$ \cite{Beringer:1900zz}.  To extrapolate to lower energy we take the approach described in the rest of this section.

Within the framework of the effective V-A interaction the infinitesimal branching ratio for the radiative muon decay $\mu^\pm \rightarrow e^\pm\,\nu\,\bar{\nu}\,\gamma$ is given by \cite{Kuno:1999jp}:
\begin{widetext}
\begin{align}
dN(e^\pm\,\nu\,\bar{\nu}\,\gamma) = \frac{\alpha}{64\pi^3}\beta\,du\frac{dy}{y}\,d\Omega_e\,d\Omega_\gamma\left[F(u,y,d) \mp\beta\vec{P}_\mu\cdot\hat{p}_e\,G(u,y,d) \mp \vec{P}_\mu\cdot\hat{p}_\gamma\,H(u,y,d) \right].
\label{dB}
\end{align} 
\end{widetext}
This is the infinitesimal probability that a muon decay produces a photon in the energy interval $dy \equiv 2\frac{dE_\gamma}{m_\mu}$ and an electron in the energy interval $du \equiv 2\frac{dE_e}{m_\mu}$ with solid angles $d\Omega_\gamma$ and $d\Omega_e$ in the \textit{muon rest frame}. Note that this notation is equivalent to that of a differential branching ratio $dB_\gamma$ as has been previously used in the literature. The photon and electron have 3 momenta $\p_\gamma$ and $\p_e$ with unit vectors expressed as $\hat{p}_\gamma$ and $\hat{p}_e$ respectively. The boost factor $\beta$ is given by $\beta = \frac{|\p_e|}{E_e}$ and the parameter $d$ is given by $1-(\beta\,\hat{p}_e\cdot\hat{p}_\gamma)$.

$\vec{P}_\mu$ is the polarisation of the muon however for this work we assume that the muons are unpolarised and hence $\vec{P}_\mu=0$. Hence, Eq.(\ref{dB}) simplifies by having terms involving the functions $G(u,y,d)$ and $H(u,y,d)$ vanish. The functional forms of $G$ and $H$ can be found in Appendix A of \cite{Kuno:1999jp}. 
$F$ is separated into factors of $r\equiv (m_e/m_\mu)^2$.
\begin{widetext}
\begin{align}
F(u,y,d)\equiv\; &F^{(0)} + rF^{(1)} + r^2F^{(3)} \;\;\;\;\text{with}:\label{F} \\
\nonumber\\
F^{(0)} = &\frac{8}{d}\lbrace y^2(3-2y)+6uy(1-y) + 2u^2(3-4y)-4u^3 \rbrace  \nonumber\\ 
&+8\lbrace -uy(3-y-y^2)-u^2(3-y-4y^2)+2u^3(1+2y) \rbrace \nonumber\\
&+2d \lbrace u^2y(6-5y-2y^2)-2u^3y(4+3y)\rbrace + 2d^2u^3y^2(2+y)\;,\\
\nonumber\\
F^{(1)} = &\frac{32}{d^2}\bigg\lbrace -\frac{y(3-2y)}{u} - (3-4y)+2u  \bigg\rbrace + \frac{8}{d}\lbrace y(6-5y) - 2u(4+y) +6u^2 \rbrace \nonumber\\
&+\lbrace u(4-3y+y^2)-3u^2(1+y)\rbrace +6du^2y(2+y)\;, \\
\nonumber\\
F^{(2)} = &\frac{32}{d^2}\bigg\lbrace\frac{(4-3y)}{u}-3\bigg\rbrace+\frac{48y}{d}\;.
\end{align}
\end{widetext}
For DM annihilation into unpolarised muons, the assumption is that one can assume an isotropic distribution of photons and marginalize over the entire solid angle of the final state photons. We also assume symmetry in the azimuthal component of the electron solid angle. Explicitly writing the solid angle differentials defined in Eq.(\ref{dB})
\begin{align*}
	d\Omega_e =& d\cos\theta_e\:d\phi_e \\
 	d\Omega_\gamma =& d\cos\theta_\gamma\:d\phi_\gamma
\end{align*} 
and integrating Eq.(\ref{dB}) over the electron azimuth and the entire photon solid angle yields the differential branching ratio
\begin{align}
	\frac{dN}{dudyd\cos\theta_e} = \frac{\alpha}				{8\pi}\cdot\beta\cdot\frac{1}{y}\cdot F(u,y,d).
\label{diffBRatio}
\end{align}
Kinematics yield the following constraints on the parameters $u, y$ and $\cos\theta_e$:
\begin{align}
&2\sqrt{r}<u<1\;\;\text{for}\;\;0<y\leq 1-\sqrt{r} \nonumber\;, \\
 &(1-y)+\frac{r}{1-y}\leq u\leq1+r\;\;\text{for}\;\;1-\sqrt{r}<y\leq 1-r\;, \nonumber\\
 & \rho\,y\,\cos\theta_e < 2(1+r) - 2(u+y)-uy \label{constraints} \;,
\end{align}  
where $\rho\equiv |\mathbf{p}_e|$ in units of $m_\mu/2$. 

\begin{figure}[t]
	\insertfig{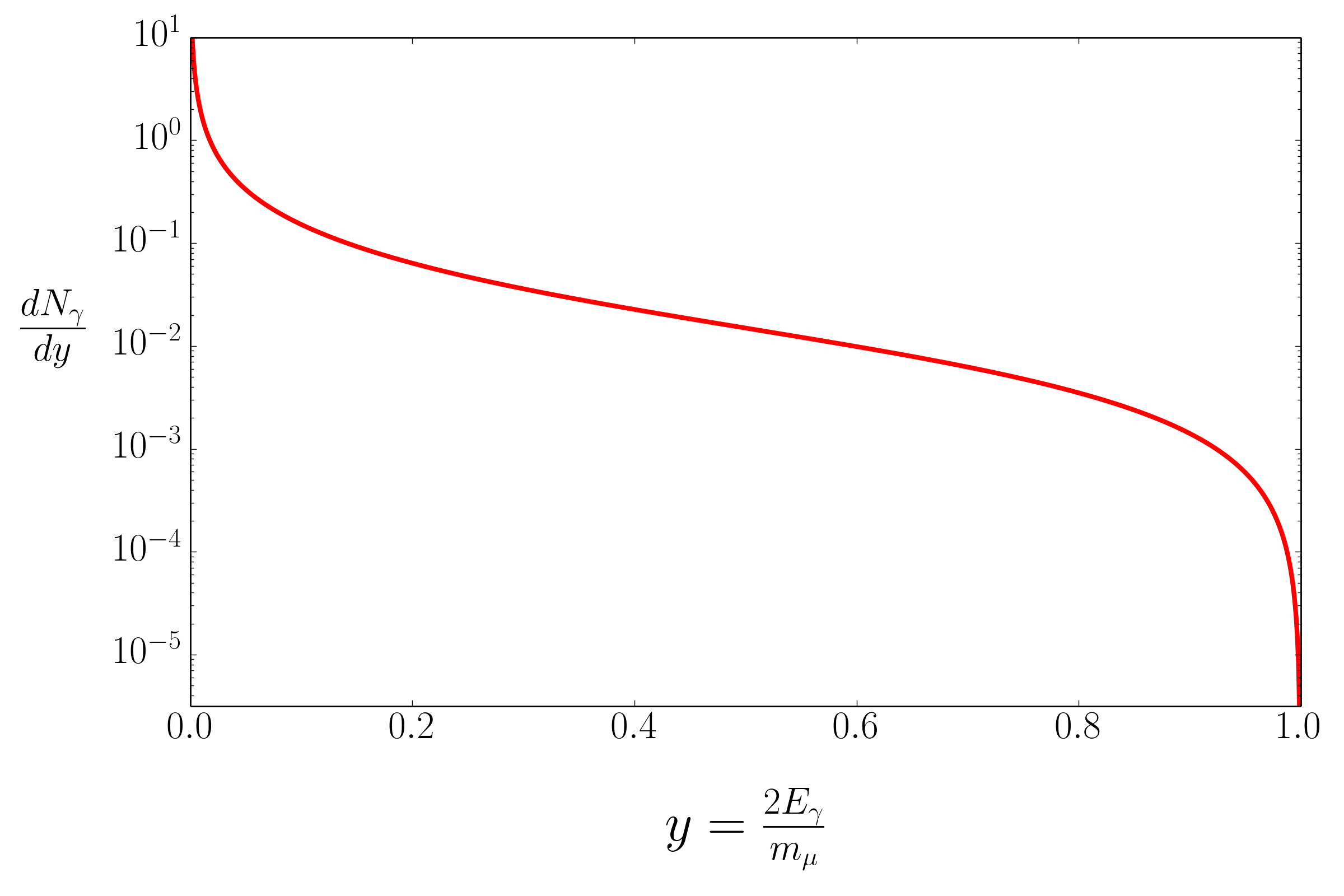}
	\caption{Differential spectrum of the $\mu^\pm\rightarrow 			  	e^\pm\bar{\nu}\nu\gamma $ decay in the muon rest frame as a function of $y=2E_\gamma /m_\mu$ after the marginalisation over electron energy and production angle. This is per DM-DM annihilation.}
	\label{fig::muRestFrame}
\end{figure}

In this paper we sample from the the probability density in Eq.(\ref{diffBRatio}) using Monte Carlo techniques subjected to the kinematic constraints in Eq.(\ref{constraints}) as detailed later in section \ref{sec::MCEventGen}. We expect the resulting spectrum of photons to be identical to what we would have obtained if we had simply integrated over the remaining degrees of freedom, namely, the electron energy and production angle.  
Upon integrating over final electron energies and polar angles ($u$ and $\theta_e$) in the limit $r\equiv\left(\frac{m_e}{m_\mu}\right)^2<<1$, the total spectrum of photons from unpolarised muons can be written:
\begin{align}
	\frac{dN_\gamma}{dy}_\text{Radiative}	= &\frac{\alpha}{3\pi}\frac{1-y}{y} \bigg(\left(3-2y+4y^2-2y^3\right)\ln\frac{1}{r}\nonumber\\
	&-\frac{17}{2} + \frac{23y}{6}-\frac{101y^2}{12} + \frac{55y^3}{12} \nonumber   \\
	 &+\left(3-2y+4y^2-2y^3\right)\ln(1-y)\bigg)
	 \label{dNdy}
\end{align}
which is shown in Fig.~\ref{fig::muRestFrame}. 
   
One immediately notices the infrared divergence present in the spectrum shown in Fig.~\ref{fig::muRestFrame}. Kuno and Okada \cite{Kuno:1999jp} mention that the singularity is resolved by corrections induced by the standard muon decay, but they do not provide details on the energy scale at which these corrections become apparent. 

We thus adopt three energy thresholds in this work at 
\begin{equation}
E_\gamma^\text{thresh} =  0.1, 1 \,\,{\rm and }\; 10 \,\,{\rm MeV}
\end{equation}
 which correspond to $y$ values of $y = $ 0.0019, 0.019 and 0.19 respectively. Of course, by defining a threshold in such a fashion, one immediately changes the total integrated branching ratio; the integral over $\frac{dN_\gamma}{dy}$ from 0 to 1 should give 1, as the function is after all a probability density. However, $\int_{y_\text{threshold}}^1 dy\; \frac{dN_\gamma}{dy} \neq 1  $. By performing a numerical integration of Eq.(\ref{dNdy}) for an energy threshold of 10 MeV ($y_0 = 0.19$), and then using a small $y$ expansion, we extrapolate the BR to lower thresholds.
\begin{align}
\label{BR}
	\text{BR}_\text{radiative}(y) \simeq B_0 + \frac{\alpha}{3\pi}\cdot\Delta(y)\;, 
\end{align}  
where $B_0 = 0.0130217$ and 
\begin{align*}
\Delta = &\left(\frac{17}{2} + 3\ln r\right)\ln\left(\frac{y}{y_0}\right) - \left(\frac{28}{3}\ + 5\ln r \right)(y-y_0) + \\
& \left(\frac{38}{8} +3\ln r\right) (y-y_0)^2  - \left(\frac{17}{6} + 2\ln r\right)(y-y_0)^3\;.
\end{align*}

\section{Derivation of the total gamma ray spectrum from muon decay}
\label{sec:spectra}
As it stands, the only photons generated in \tt PYTHIA \rm that arise from processes that produce muons in the final state come from FSR. To make this clear, the final state here refers to a muon in the final state of the dark matter annihilation process, \textit{not} the muon decay itself, which occurs once the muon has propagated. That is, the radiative decay shown in Fig. \ref{fig::RadDecay} is treated as a completely different process to FSR. Thus, there is no overlap (and hence no double counting of photons) between the two processes. The spectrum of photons in the muon rest frame arising from FSR for the process $\chi\chi\rightarrow \mu^-\mu^+$ is given by equation 4 of \cite{Essig:2009jx}:
\begin{align}
\label{FSRSpect}
	\frac{dN_\gamma}{dy}_\text{FSR} = \frac{\alpha}{\pi}\left(\frac{1+(1-y)^2}{y}\right)\left(\ln\left(\frac{s(1-y)}{m_f^2}\right) - 1\right)\;.
\end{align}
The total spectrum of photons in the muon rest frame is then the sum of radiative and FSR contributions given by Eq.(\ref{dNdy}) and Eq.(\ref{FSRSpect}) respectively
\begin{align}
\label{dNdyTot}
	\frac{dN_\gamma}{dy} = \frac{dN_\gamma}{dy}_\text{Rad} + \frac{dN_\gamma}{dy}_\text{FSR}
\end{align}

For the primary case of DM annihilation directly to muons $\chi \chi \rightarrow \mu^+\mu^-$ one can write down the spectrum of photons in the DM annihilation frame using the good approximation from \cite{Essig:2009jx}:
\begin{align}
	 \frac{dN_\gamma}{dx} = 2\,\int^1_x\;dy\,\frac{1}{y}\frac{dN_\gamma}{dy}\;, \label{2muSpec}
\end{align}
where $x\equiv \frac{E_\gamma}{m_\chi}$ and $\frac{dN_\gamma}{dy}$ is the muon rest frame spectrum from Eq.(\ref{dNdy}).  

For the case of a mediated annihilation to muons $\chi\chi\rightarrow \phi\phi\rightarrow \mu^+\mu^-\mu^+\mu^-$ there needs to be two boosts: first from the muon rest frame to the $\phi$ frame, then from the $\phi$ frame to the DM frame. Let $E_\phi$ be the energy of the photon in the $\phi$ rest frame and $\omega = \frac{2E_\phi}{m_\phi}$. Then,   
\begin{align}
	\frac{dN_\gamma}{d\omega} = 2\,\int^{\text{min}(1,\frac{2y}{1-\beta_1})}_{\frac{2y}{1+\beta_1}}dy\,\frac{1}{y}\frac{dN_\gamma}{dy}\;,
	\label{dNdw}
\end{align}     
 where $\beta_1 = \sqrt{1-\frac{4m_\mu^2}{m_\phi^2}}$, takes us from the muon rest frame to the $\phi$ rest frame.  
The next boost with $\beta_2 = \sqrt{1 - \frac{m_\phi^2}{m_\chi^2}}$ takes us to the DM annihilation frame. The form of the spectrum in the DM annihilation frame $\frac{dN_\gamma}{dx}$ can be simplified slightly and be written in the form
\begin{align}
	\frac{dN_\gamma}{dx} = \frac{1}{\beta_2}\big[I(\omega_\text{min}) - I(\omega_\text{max})\big],\label{4muSpec}
\end{align}  
where 
\begin{align}
	I(\omega_i) = \int^\infty_{x_i}\, d\omega\,\frac{1}{\omega}\,\frac{dN_\gamma}{d\omega}
\end{align}
and $\omega_\text{max} = 2x(1+\beta)\gamma $, $\omega_\text{min} = {2 x}/{(1+\beta)}$.

It should be noted that in the limit $m_\phi<<m_\chi$, $\omega_\text{min} \rightarrow x$ and $\omega_\text{max}>> 1$. This then leads to the expression seen in various parts of the literature, for example \cite{Essig:2009jx}
\begin{align}
	\frac{dN_\gamma}{dx} = 2\,\int^1_x\;dy\,\frac{1}{\omega}\frac{dN_\gamma}{d\omega}\;,
\end{align}
which assumes the $m_\phi << m_\chi$ limit. This assumption was not deemed valid for this work due to the inclusion of heavy mediators in the model parameter space.

\subsection{Monte Carlo Event Generation}
\label{sec::MCEventGen}
   
   \label{sec:monteCarlo}
\begin{figure}[t]
	\centering
	\includegraphics[scale=0.5]{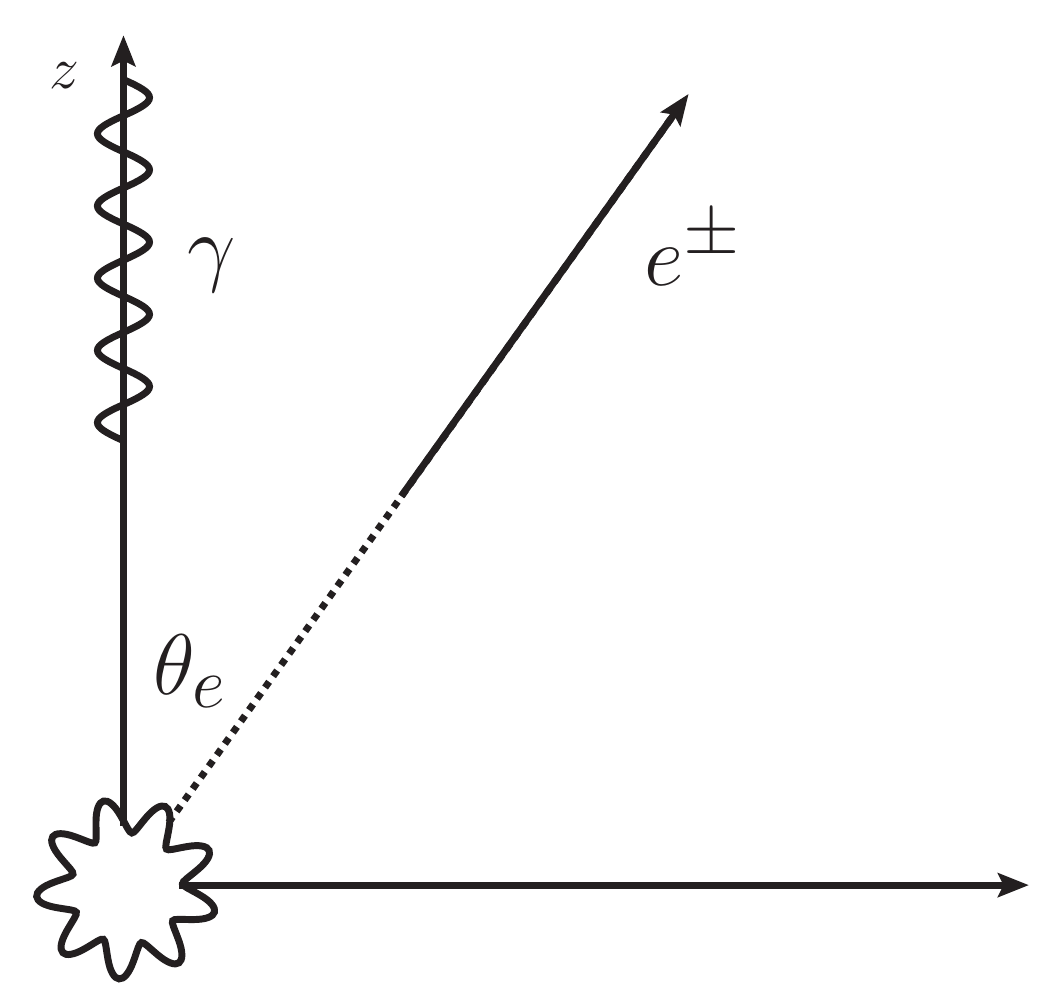}
	\caption{Radiative muon decay in the muon rest frame as set up in \tt PYTHIA \rm. Shown is a radiative event in which the muon decays into a photon defined along a $z$ axis with energy $y = \frac{2E_\gamma}{m_\mu}$ and an electron is emitted at an angle $\theta_e$ from this axis with energy $x=\frac{2E_\gamma}{m_\mu}$. The three parameters $(x,y,\theta_e)$ are sampled from Eq.(\ref{diffBRatio}) after the marginalization of all other energy and angular variables. }
	\label{fig::restFrameDecay}
\end{figure} 
\begin{figure*}[ht]
	\makebox[\linewidth][c]{%
\insertdoublefig{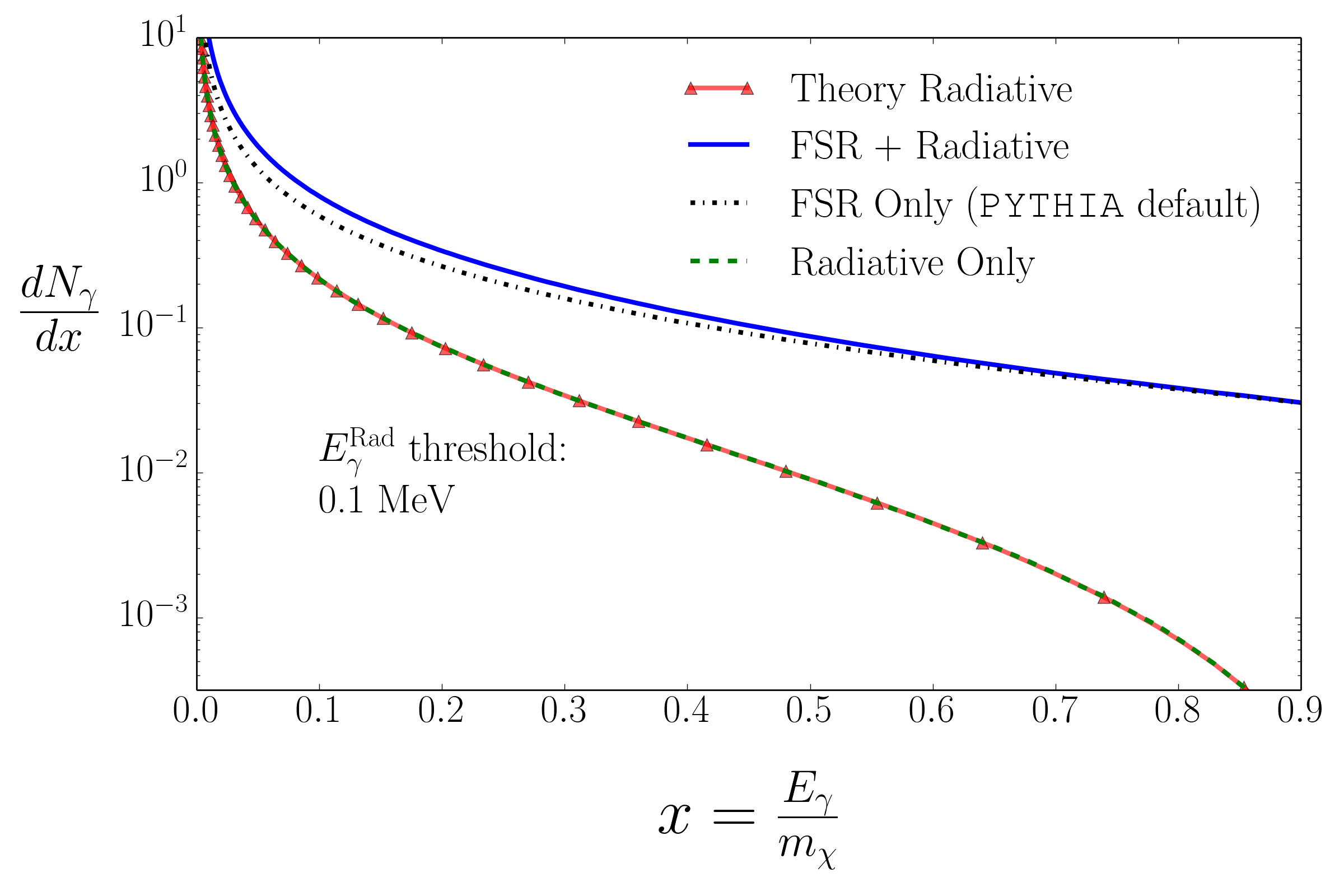}{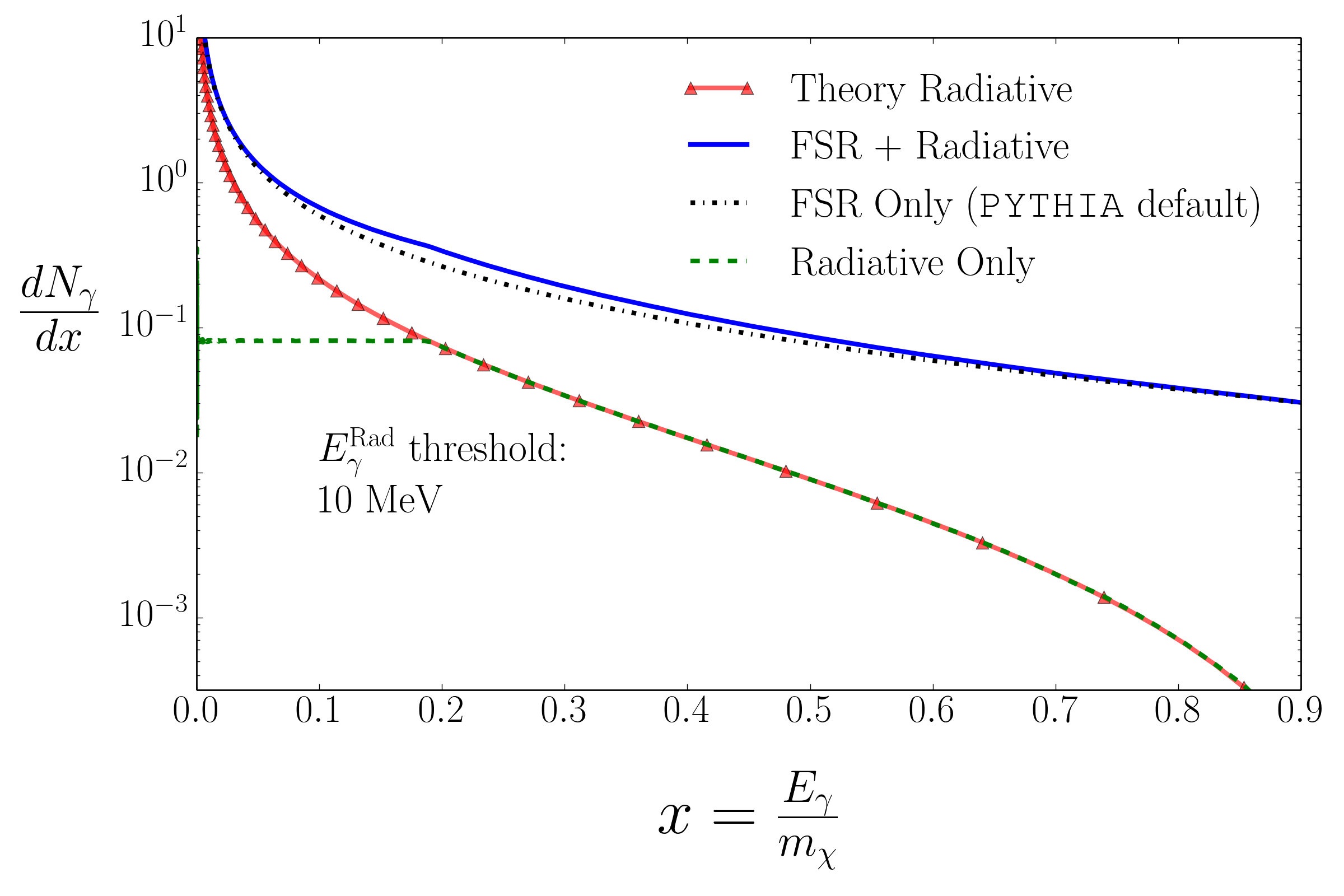}
}
	\caption{Spectrum of photons per $\chi\chi\rightarrow\mu\mu$ annihilation in the DM annihilation rest frame for $m_\chi = 100$ GeV. Shown in (a), (b) are the two extreme scenarios invoking a 0.1 and 10 MeV radiative photon energy threshold respectively. The black and green curves are the radiative and FSR only spectra respectively, while the blue curve is the sum of the two contributions.  
Also shown for validation is the radiative contribution obtained from theory. This curve is evaluated by first taking the muon rest frame spectrum in Eq.(\ref{dNdy}) and boosting with Eq.(\ref{2muSpec}) to get to the DM rest frame.}\label{fig::validation}
\end{figure*}

We use the  MC particle event generator \tt PYTHIA 8.175 \rm to simulate $10^9$ muon decays arising from DM annihilation as follows. Since \tt PYTHIA \rm cannot directly simulate DM annihilation we follow the procedure usually undertaken in this situation which involves first    
producing an artificial resonance with the exact four momentum characteristics of two DM particles which subsequently decays to muons $e^-\:e^+\rightarrow \psi\rightarrow \mu^+ \mu^-$. This gives the spectrum of photons that would originate from $\chi\chi \rightarrow \mu^+ \mu^-$ in the DM rest frame. We then obtain the spectrum of photons for the four muon final state  $\chi\chi\rightarrow \phi \phi\rightarrow 4\mu$ in the DM annihilation rest frame by taking this resultant spectrum and then boosting by Eq.(\ref{4muSpec}) to get to the DM rest frame.

To correctly generate a spectrum of photons from the radiative muon decay consistent with the spectrum in Eq.(\ref{dB}) we implement an acceptance and rejection Monte Carlo technique to sample from the probability distribution given in Eq.(\ref{diffBRatio}). 
For each of the specified lower thresholds (0.1, 1 and 10 MeV) we use Eq.(\ref{BR}) to calculate the BR of the radiative decay. 
We then generate a random number $R\in[0,1]$. If $$R < \text{BR}_\text{radiative}$$ then \tt PYTHIA \rm calls our sampling routine to perform a radiative decay. If not, \tt PYTHIA \rm will proceed to perform the             
 dominant 3-body Michel decay already available through its standard                   mechanisms. The result after the generation of a large enough event sample is a population of photons from muon decays that reflect this initial BR. In order to generate these photons we sample uniformly from the differential branching ratio shown in Eq.(\ref{diffBRatio}). To do so we first set up the muon decay as represented in Fig.~\ref{fig::restFrameDecay}: define a $z$ axis to be the direction of the emitted photon with energy $E_\gamma$. Then, define an electron emitted at some angle $\theta_e$ from this axis to have energy $E_e$ and 3 momentum $\p_e$. Neutrinos are generated isotropically in their rest frame and boosted to the muon rest frame whilst ensuring energy and momentum conservation.

\begin{table*}[t]
\centering
\makebox[\linewidth][c]{%
\begin{tabular}{c c }
Process &  \texttt{PYTHIA 8} option \\ 
\hline 
Turn off initial state QED radiation & \texttt{ pythia.readString("PartonLevel:ISR = off")}\\ 

Turn on FSR for $E_\text{CM}>20$ GeV  &\texttt{pythia.readString("PartonLevel:FSR = on")} \\ 

Turn on FSR for $E_\text{CM}<20$ GeV & 
\texttt{ pythia.readString("ParticleDecays:allowPhotonRadiation = on")}\\ 
\hline
\end{tabular} 
}
\caption{The relevant \texttt{PYTHIA} options for setting the correct FSR parameters which are implemented where one would initialize the main process. } 
\label{table::pythiaSettings}
\end{table*}

To increase efficiency we use an envelope function in the accept/reject method that encompasses Eq.(\ref{diffBRatio}) on as much of the domain space ($x$ ,$y$  ,$\cos\theta_e$) as possible.  The functional form of the envelope is 
\begin{align}
	F_\text{envelope}(x,y) = 24.83\:\frac{(x+0.175)^2}{y}\;.
\end{align}  
Energies $x$ and $y$ are sampled uniformly from this envelope distribution, whilst $\cos\theta_e$ is sampled over the interval $[-1,1]$. We then impose the kinematic constraints shown in Eq.(\ref{constraints}) on the three parameters. If and only if the kinematic constraints are satisfied for every parameter ($x$, $y$, $\cos\theta_e$) in the iteration do the parameters get allowed to partake in the acceptance and rejection. 
The number $u$ is randomly generated on the interval $[0,1]$. If 
\begin{align}
	u\,F_\text{envelope}(x,y) \leq \frac{dB}{dxdyd\cos\theta_e}(x,y,\cos\theta_e)
\end{align}   
then the event is accepted, otherwise, the process is reiterated. All events are binned corresponding to their energy. The differential energy spectrum of photons is then obtained by normalising by the total number of events. This is effectively the energy spectrum per DM annihilation.  
In Fig. \ref{fig::validation} we show extent of the radiative correction for the spectrum of photons arising from a 100 GeV DM annihilation into two muons $\chi\chi\rightarrow\mu^+\mu^-$ in the DM rest frame for the two cases of a 0.1 and 10 MeV threshold. Shown are the spectra from the radiative decay only, FSR only and FSR + radiative contributions. For validation we also show the analytical curve which has been obtained by using Eq.(\ref{2muSpec}) on Eq.(\ref{dNdy}).   
We see that there is a noticeable deviation from the FSR only contribution at soft photon energies arising from the radiative correction. As expected, this deviation is most prominent  for the lowest energy threshold of 0.1 MeV, since in this case, we are approaching the infared divergence seen in Fig.~\ref{fig::muRestFrame}, and hence expect more soft photons in the spectrum.    


\subsection{\texttt{PYTHIA 8} settings}

\texttt{PYTHIA} has settings for initial state and final state QED radiation. In this work, we switch any initial state radiation merging from the artificial $e^+ e^-$ resonance off. By default, FSR is switched on for leptonic processes, but only for cases with $E_\text{CM} > 20$ GeV. For final state QED radiation with $E_\text{CM} \leq 20$ GeV, we use another option called ``allowPhotonRadiation". The spectra from this option at 20 GeV matches that of the default FSR, except                                   
for photons below $\sim$ 0.1 MeV. This in part was the motivation behind adopting the minimum threshold mentioned previously. We include the relevant \texttt{PYHTIA} options used to implement the correct FSR for maximum reproducibility in table \ref{table::pythiaSettings}.
For more detail on these options, see the particle decay section of the \texttt{PYTHIA} manual \cite{sjostrand2008jhep05}.

We also provide users of \texttt{micrOMEGAs} with updated look up tables that contain photon spectra including the radiative correction for a variety of WIMP masses (see appendix \ref{sec:micro}). These are formatted to replace the default spectra which only include an FSR component.

\begin{figure*}[t]
\insertdoublefig{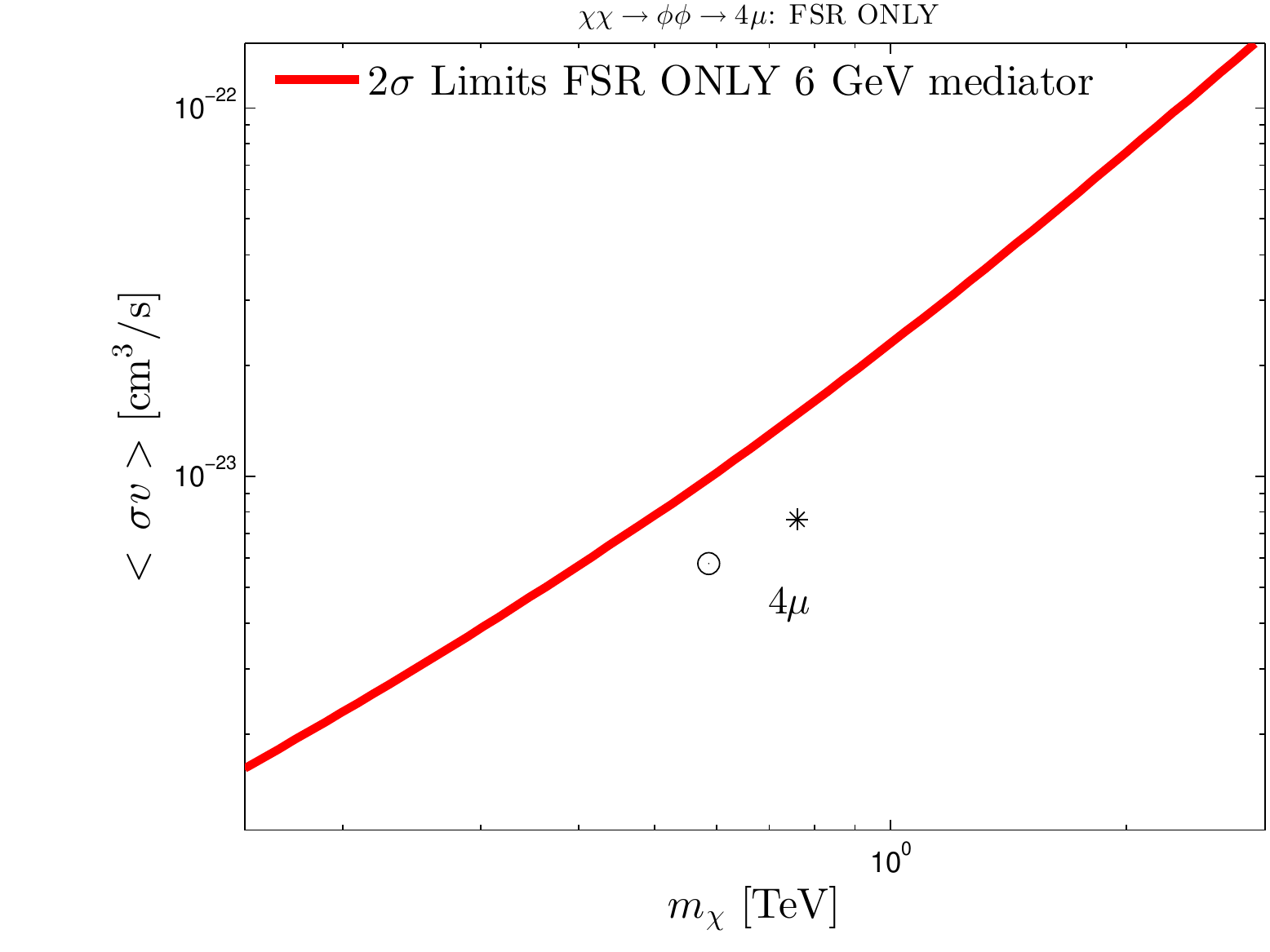}{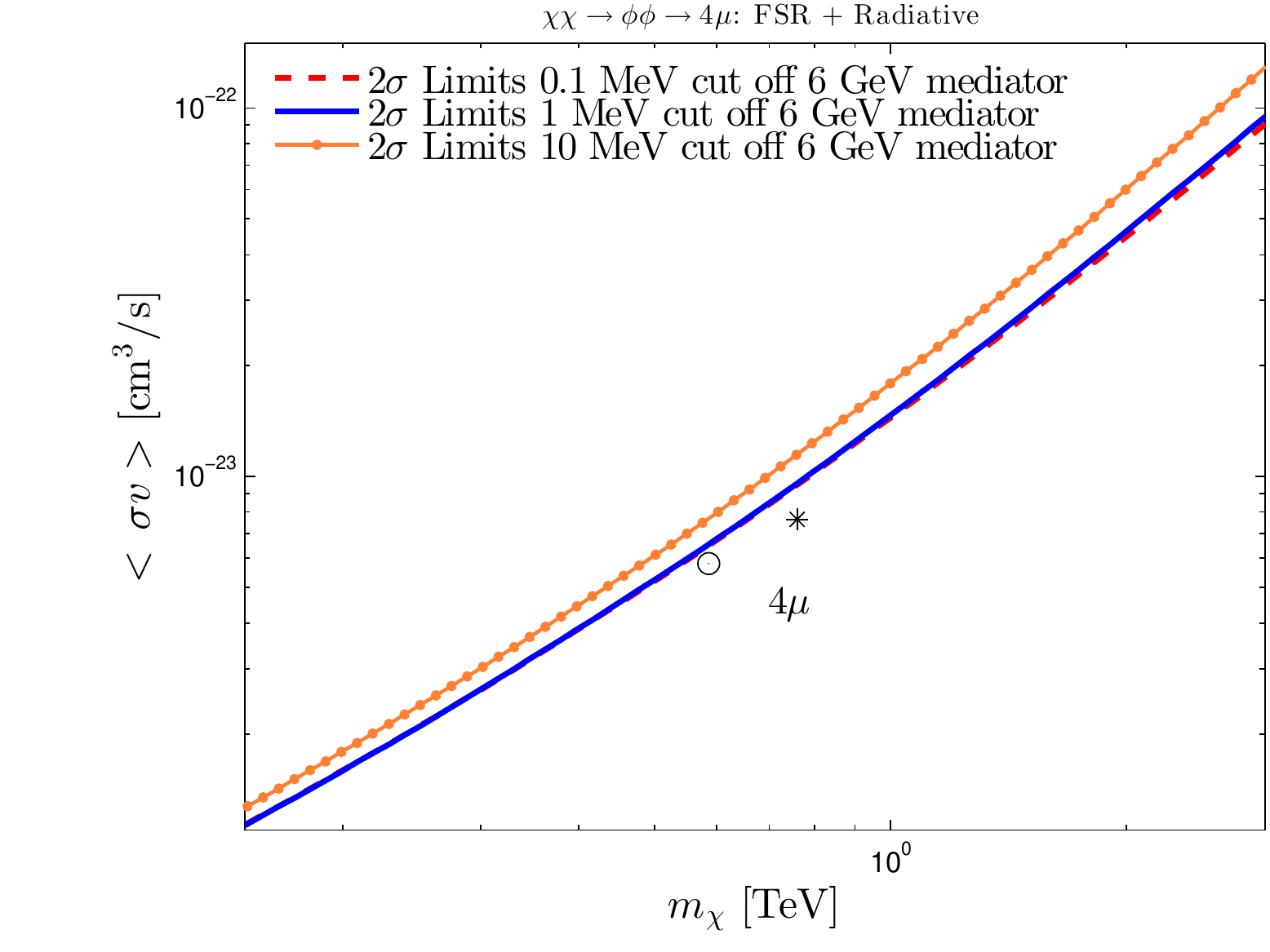}
\caption{The AMS-02 best fit parameters (circle and star) and \texttt{Pass 8} Fermi-LAT dwarf spheroidal upper bounds (curves) on $\langle \sigma v\rangle$ and $m_\chi$ with and without the radiative correction to the decay of the muon arising from $\chi\chi\rightarrow\phi\phi\rightarrow 4\mu$ with $m_\phi = 6$ GeV. In both figures, the circle corresponds to the MED propagation parameters, and the star corresponds to the `best fit' propagation parameters.
Left: The limit is presented with no radiative correction included. In this case, the AMS-02 interperetation still survives. Right: FSR + radiative correction included. The red, blue and green curves correspond to the three low energy thresholds of 0.1, 1 and 10 MeV considered in this work. The strongest limit is provided when radiative photons of energies down to 0.1 MeV are allowed.   }\label{fig::FermiDwarfLimits}
\end{figure*}

\section{Constraining the final AMS-02 explanation}
\label{sec:ams}

Boudaud et al. \cite{Boudaud:2014dta} use the benchmark set of CR propagation parameters known as ``MED" to show that the DM annihilation channel $\chi\chi\rightarrow\phi\phi\rightarrow\mu^+\mu^-\mu^+\mu^-$ (where the muons promptly decay to $e^\pm$) provides a good fit to the steady increase of the positron spectrum observed by AMS-02 \cite{Accardo:2014lma}. Furthermore, by accounting for the systematic uncertainties on the CR parameters, they perform a scan over the allowed parameter space to obtain a `best fit' set of propagation parameters. These fits are shown as the circle (MED) and star (best-fit) in Fig. \ref{fig::FermiDwarfLimits}.

DM annihilations that produce $e^\pm$, either directly or through decays and showering of the primary annihilation products, will invariably also produce gamma rays. One can then constrain the AMS-02 fits mentioned above with gamma-ray observations. 
The Fermi dwarfs are some of the most DM rich objects known to astrophysics. Given their relatively close proximity they thus make excellent targets for indirect detection. Fermi-LAT has surveyed 25 of these dwarfs in the energy range 500 MeV to 500 GeV in search of a significant excess of $\gamma$ rays to attribute to DM annihilation. At present, no such excess has been detected.
As previously mentioned, the authors of \cite{Lopez:2015uma} have excluded all potential final states using the \texttt{Pass 7} Fermi-LAT dwarf data except for the four muon final state $\chi\chi\rightarrow \phi\phi\rightarrow 4\mu$ at greater than the $2\sigma$ level. 

In this section, we use the new Fermi-LAT likelihoods from the \texttt{Pass 8} event level analysis \cite{Ackermann:2015zua} to perform a combined dwarf likelihood analysis to constrain the parameter space of DM annihilation into the four muon final state. For each dwarf and energy bin indexed $k,j$ respectively Fermi provides a likelihood $\mathcal{L}_{k,j}$ for a corresponding energy flux
\begin{align}
	s_{k,j} = \int^{E_{j,\text{max}}}_{E_{j,\text{min}}}\: E \,\frac{d\phi_k}{dE}dE
\end{align}
where $\frac{d\phi_k}{dE}$ is the differential gamma ray flux from the spherodial $k$. The differential flux is a function of the spectrum of photons from the annihilation process of interest $\frac{dN_\gamma}{dE} = \frac{1}{m_\chi}\frac{dN_\gamma}{d x}$. In this case, this is the photon spectrum resulting from $\chi\chi\rightarrow\phi\phi\rightarrow 4\mu$ which is generated by boosting the muon rest frame number spectrum obtained by Monte Carlo using Eq.(\ref{dNdw}) followed by Eq.(\ref{4muSpec}).
To firstly illustrate the effects of including the effects of radiative muon decay, we show
limits with and without our radiative correction implemented at $2\sigma$ (95.45\% C.L) in Fig. \ref{fig::FermiDwarfLimits} for a mediator mass of $m_\phi=6$ GeV.
As shown in the left panel, we find that FSR alone does not produce a limit that excludes the AMS-02 explanation for a 6 GeV mediator at greater than the 2$\sigma$ level. However, as shown in the right panel, the addition of the radiative component makes the bounds tighter and begins to create tension with the best fit  AMS-02 parameters for MED propagation parameters. We see that the strongest constraint comes from allowing radiative photons down to the lowest energy threshold considered of 0.1 MeV, which is expected since we would expect to see more photons arising from the steepening of the spectrum at lower energies. 
%
\begin{figure*}[t]
\insertdoublefig{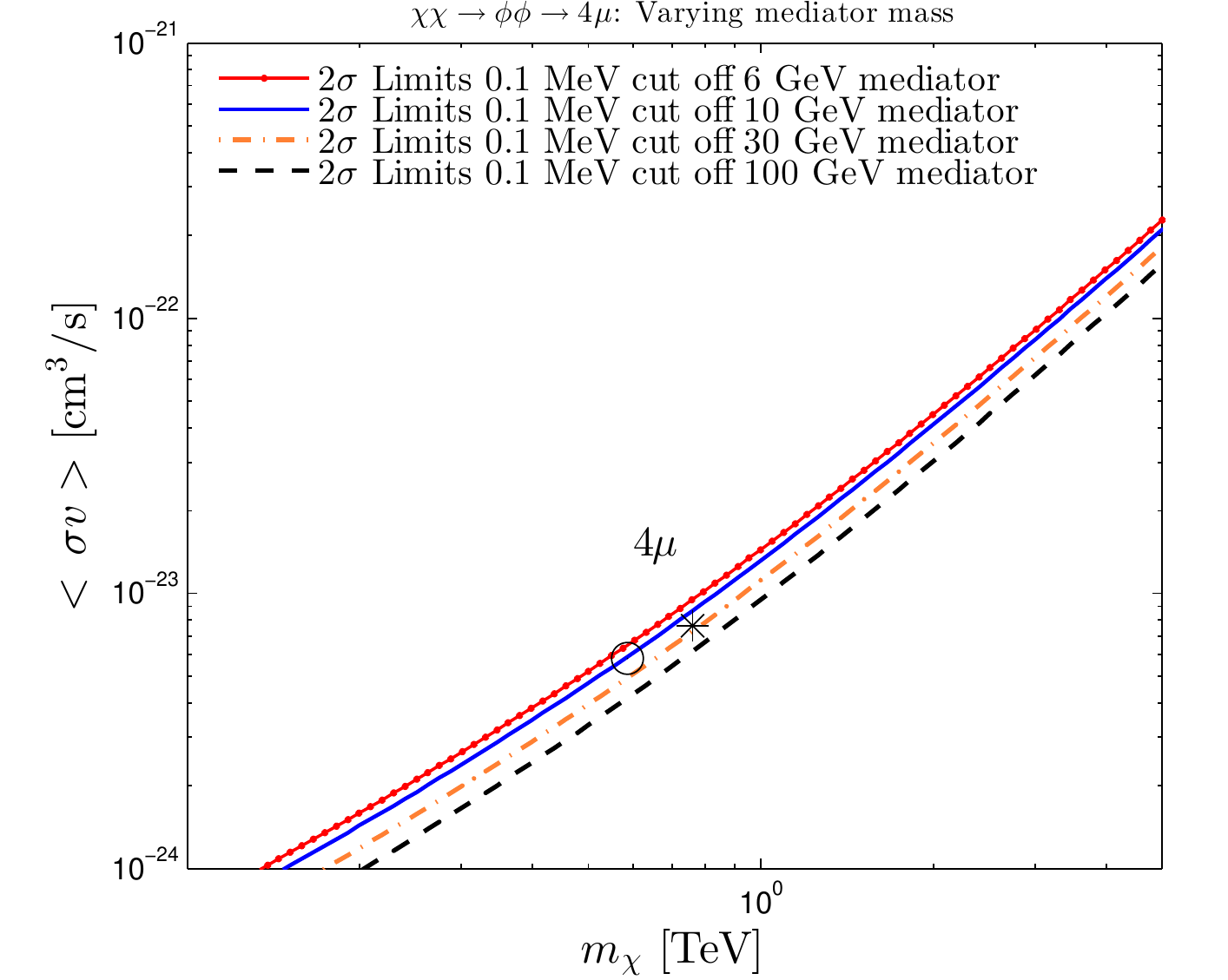}{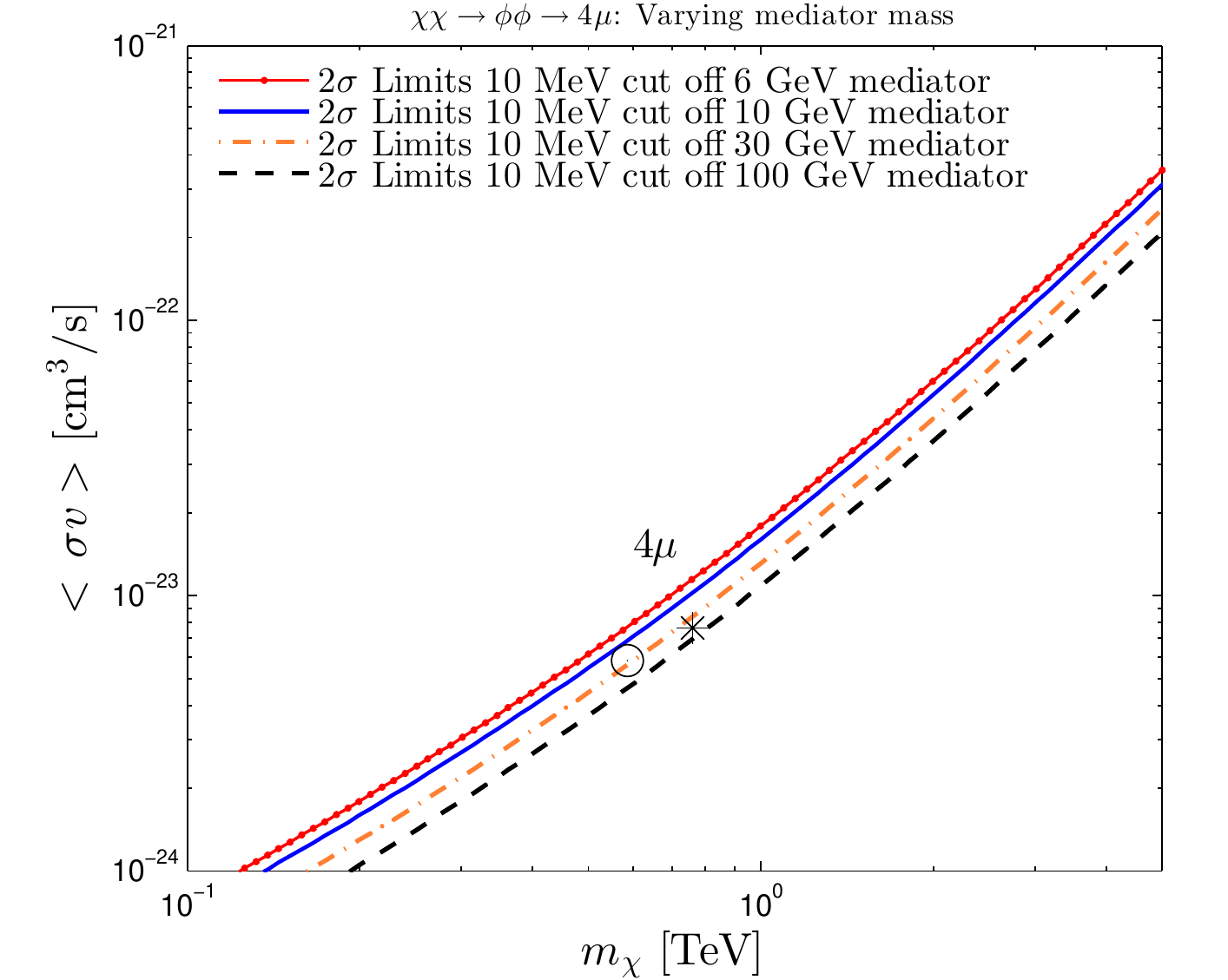}
\caption{Curves show Fermi-LAT upper bounds on the $\chi\chi\rightarrow \phi\phi\rightarrow 4\mu$ channel at varying $m_\phi$ with a radiative photon threshold of 0.1 MeV (left) and 10 MeV (right). Circle and star indicate AMS-02 best fit parameters as described in the previous figure.}
\label{fig::FermiDwarfLimitsMediator}
\end{figure*}
We show the effect of varying the mediator mass on the Fermi-LAT limit in Fig. \ref{fig::FermiDwarfLimitsMediator}. We display $2\sigma$ upper limits for mediator masses of 6,10, 30 and 100 GeV. Here we show the limits using a 0.1 MeV and a 10 MeV threshold. The limits become stronger as the mass of the mediator $\phi$ increases, excluding at greater than the $2\sigma$ level the best-fit AMS-02 explanation for mediators with $m_\phi> 100$ GeV  for the conservative case of a 10 MeV threshold and $m_\phi>30$ GeV for the 0.1 MeV case. The MED fit only survives at $2\sigma$ for $m_\phi>10 $ GeV for the 10 MeV threshold while it is excluded  by all but $m_\phi=6$ GeV in the 0.1 MeV case. 
We choose to stop at 100 GeV since we see that the AMS best-fit and MED points are excluded at greater than the $2\sigma$ level even in the most conservative case of a 10 MeV threshold.

\section{Fitting the Fermi GC excess}
\label{sec:fermi}

\begin{figure}[t]
\insertfig{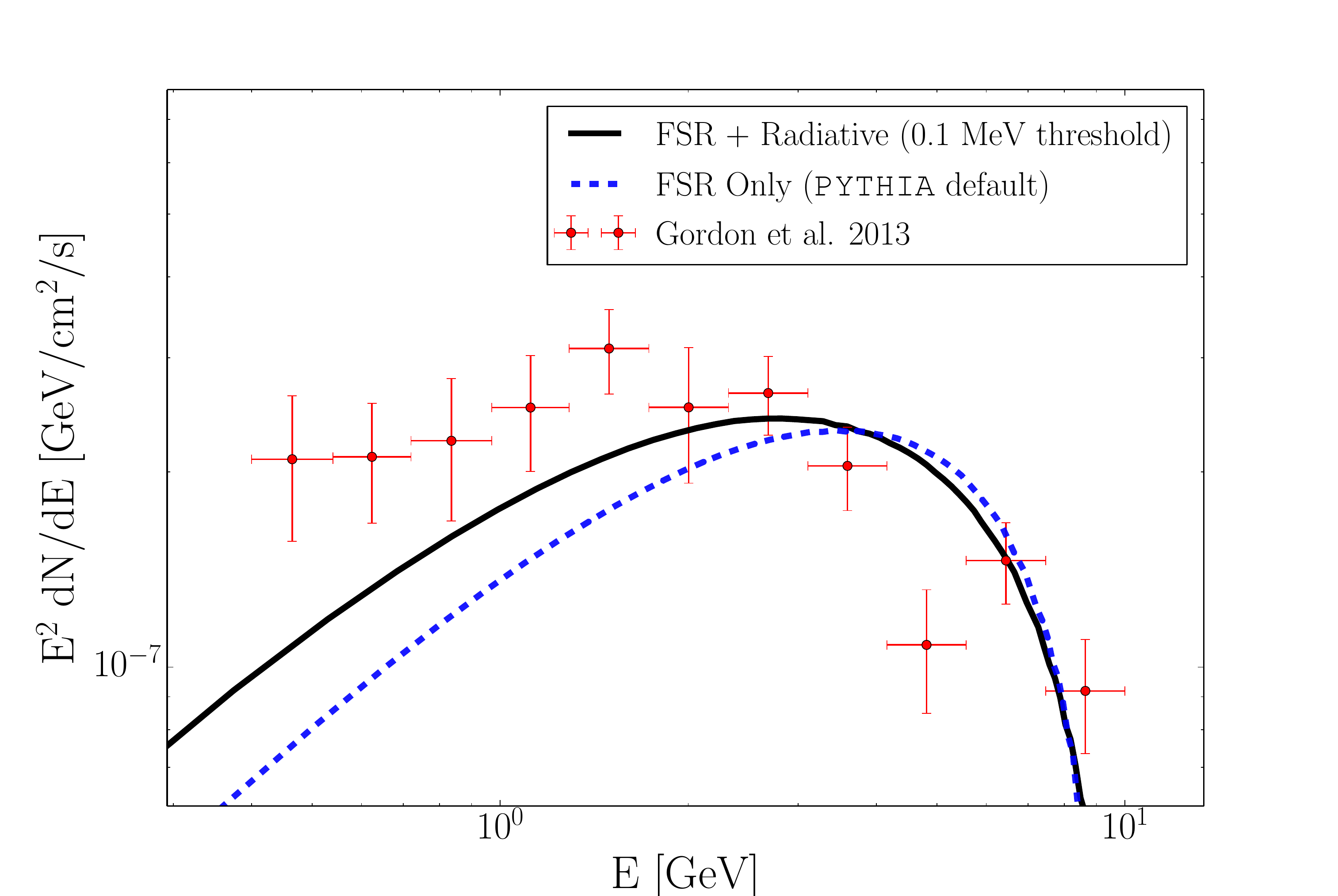}
\caption{The $7^\circ\times7^\circ$ extended residual presented by \cite{Gordon:2013vta} corresponding to a best fit NFW template with an inner slope of $\gamma=1.2$ from the GC is shown with the red crosses. The vertical error bars show show the maximum of the $1\sigma$ systematic and statistical error. The spectral point $E^2dN/dE$ is evaluated at the logarithmic midpoint of horizontal bars. The best fit spectrum of FSR photons originating from the $\chi\chi\rightarrow \phi\phi\rightarrow \mu^+\mu^-\mu^+\mu^-$ annihilation is shown with the blue dahsed line, but it is not at all a good fit to the data. Best fit parameters for the FSR only case are $m_\chi =$ 9.87 GeV and $\langle \sigma v\rangle = 8.68\times 10^{-27}$ cm$^3$/s.  We show the best fit radiative correction to the annihilation spectrum with the solid black curve. This comes from using a lower energy threshold of 0.1 MeV and has best fit parameters $m_\chi =$ 10.34 GeV and $\langle \sigma v\rangle = 7.10\times 10^{-27}$ cm$^3$/s.}\label{fig::GCFit}
\end{figure}

In this section we show that including the radiative correction to the muon decay spectrum originating from the $\chi\chi\rightarrow\phi\phi\rightarrow 4\mu$ process considered earlier can improve the fit to the gamma-ray excess seen in a $7^\circ\times7^\circ$ region centered on $(l,b) = (359.9442^\circ,-0.0462^\circ)$ as presented in \cite{Gordon:2013vta}.
The prompt differential gamma-ray flux [GeV$^{-1}$ cm$^{-2}$ s$^{-1}$]  arising from WIMP annihilation into a four muon final state within an angular field of view centred on a galactic latitude $b$ and longitude $l$ along a line of sight $s$ is

\begin{align*}
\frac{d\phi}{dE} = \left(\frac{1}{8\pi}\frac{\langle\sigma v\rangle}{m_\chi^2}\,\frac{dN_\gamma}{dE}\bigg|_{
\chi\chi\rightarrow\phi\phi\rightarrow 4\mu}
   \right)\,\int\limits_{\text{f.o.v}}\;\int\limits_\text{l.o.s}\,ds \;d\Omega\; {\rho^2(r)}\;,
\end{align*}
where $r=\sqrt{R_\odot-2sR_\odot\,\cos(b)\cos(l)+s^2}$, ${dN_\gamma}/{dE}$ is the total number spectrum of photons per DM annihilation from Eq.(\ref{dNdyTot}), $d\Omega = db\,dl\cos (b)$ and we take $R_\odot = 8.25$ kpc as the distance from the sun to the GC. We assume a generalized Navarro-Frenk-White (NFW) halo profile \cite{0004-637X-490-2-493,2002ApJ...573..597K}


\begin{align*}
\rho(r) = \rho_\odot \left(\frac{r}{R_\odot}\right)^{-\gamma}\left(
\frac{1+\left(\frac{r}{R_s}\right)^\alpha}{1+\left(\frac{R_\odot}{R_s}\right)^\alpha}
\right)^{-(\beta-\gamma)/\alpha}\;,
\end{align*}
with $R_s = 23.1$ kpc, $\alpha=1$, $\beta =3$. $\gamma=1.2$ and $\rho_\odot=0.36$ GeV cm$^{-3}$.  
In Fig. \ref{fig::GCFit} we fit the GC excess with spectra originating from prompt FSR only as well as with the radiative correction implemented.
We calculate a test statistic which we assume follows a $\chi^2$ distribution with $11-2 = 9$ d.o.f (11 data and 2 physical parameters $\langle \sigma v\rangle$ and $m_\chi$). This is given by
\begin{align*}
	\chi^2 = \sum_i\frac{\left(F_i^\text{Data}-F_i\right)^2}{\sigma^2_{i\:\text{stat}}+\sigma^2_{i\:\text{sys}}}\;,
\end{align*} 
where we adopt the convention $F = E^2\frac{d\phi}{dE}$ and the index $i$ runs over the 11 data bands with ($1\sigma$) statistical and systematic errors tabulated in table V of \cite{Gordon:2013vta}. We define a `good fit' to be that which gives $\chi^2_\text{min}<27.88$ which corresponds to a p-value of $p>10^{-3}$.
The FSR-only curve does not provide a good fit with best fit parameters $m_\chi= 9.870$ GeV and $\langle \sigma v\rangle = 8.860\times10^{-27}$ cm$^3\,$s$^{-1}$ and a minimum $\chi^2_\text{min} = 39.70$. 
The extra soft component of photons induced by the radiative correction gives rise to a slight, but significant widening of the curve which is maximised for light mediators. Since \texttt{PYTHIA} can only handle mediators down to 3 GeV, we adopt this value for the mass of $\phi$. We find that the best fit parameters are for the 0.1 MeV radiative threshold with a DM mass of 10.33 GeV, $\langle \sigma v\rangle = 7.101\times10^{-27}$ cm$^3\,$s$^{-1}$ and a minimum $\chi^2_\text{min} = 24.26$ which corresponds to a `good' fit. A better fit is of course expected due to the larger population soft photons from invoking the radiative correction. We provide a table summarising the best fit parameters for the other low energy thresholds in table \ref{table::scanParams}. We display for each threshold scenario the $\chi^2_\text{min}/\text{d.o.f}$ to illustrate the significance of adding the radiative component compared to the FSR only case.

\begin{table}[h]

\begin{tabular}{c c c c}
Energy threshold &  $\langle \sigma v\rangle$ [cm$^3$/s] &  $m_\chi$ [GeV] & $\chi^2_\text{min}$/d.o.f \\ 
\hline 
FSR Only & $8.68\times 10^{-27}$ & 9.87 & 4.41 \\ 

0.1 MeV & $7.10\times 10^{-27}$ & 10.40 & 2.70 \\ 


10 MeV & $6.94\times 10^{-27}$ & 9.95  & 3.43 \\ 
\hline
\end{tabular} 
\caption{Best fit parameters to the GC excess the varying radiative photon energy thresholds. Shown are results for the two extreme threshold scenarios of 0.1 and 10 MeV. Also included are the best fit parameters for the FSR only component.}  \label{table::scanParams}
\end{table}


%
%
%

\section{Conclusions}
\label{sec:conclusions}
The particle nature of DM is still unknown. If the DM is a WIMP, the astrophysical search for products of WIMP annihilation remains a viable means of detection.
In this paper, we looked further at the case of gamma rays produced via dark matter annihilation into muonic final states. In particular we showed that including the extra photons that arise from the radiative decay of the muon proves to be significant, specifically when applied to the recent electron-positron anomaly identified by AMS-02 and the Fermi-LAT galactic center excess. Indeed, the radiative correction we have generated is applicable to any model with a muonic final state.

After a review on the background theory of radiative muon decay we computed and presented the revised gamma ray spectra that arises after adding the radiative correction to the already \texttt{PYTHIA} default FSR component. We show that the radiative correction significantly increases the population of soft photons arising from muon decays. Due to an infared singularity in the radiative photon spectrum, we adopt three lower thresholds of 0.1, 1 and 10 MeV. As expected we saw that the enhanced population of soft photons arising from muon decays is sensitive to this threshold. 

To demonstrate the significance of including the radiative correction to the muon decay, we applied our results to the electron-positron anomaly seen in AMS-02. Previous studies have shown that the four muon final state is the only one not excluded by the $\texttt{Pass 7}$ Fermi-LAT dwarf constraints on gamma ray emission at the $2\sigma$ level. We calculated the $2\sigma$ upper limits for the recent $\texttt{Pass 8}$ event level analysis for both FSR only and FSR + Radiative contributions arising from $\chi\chi\rightarrow \phi\phi \rightarrow4\mu$. We saw that including the radiative correction strengthens the limits and excludes (at the $2\sigma$ level) the `best-fit' AMS-02 explanation for mediators with $m_\phi> 100$ GeV  for the conservative case of a 30 MeV threshold and $m_\phi>10$ GeV for the 0.1 MeV case. The MED fit only survives for $m_\phi>10 $ GeV for the 10 MeV and only for $m_\phi>6$ GeV in the 0.1 MeV case. 


We showed that the total prompt photon spectrum (FSR + radiative) arising from the $\chi\chi\rightarrow\phi\phi\rightarrow 4\mu$ annihilation provides a significantly better fit to the $7^\circ\times7^\circ$ gamma-ray residual observed by \cite{Gordon:2013vta} at the galactic center than when FSR contributions only are considered. We perform a least-squares goodness of fit test over 11 data and 2 physical parameters, namely the velocity averaged cross-section $\langle\sigma v\rangle$ and the DM mass $m_\chi$. Given our choice of p value  $p=10^{-3}$, we found that the radiative spectrum with the lowest threshold (0.1 MeV) provided a good fit with best fit parameters $m_\chi=$ 10.34 GeV, $\langle \sigma v\rangle = 7.10\times10^{-27}$ cm$^3\,$s$^{-1}$. The FSR only spectrum did not yield a good fit.

The software $\texttt{micrOMEGAs}$ uses look up tables to generate spectra of final states from user input model parameters.  We provide users of \texttt{micrOMEGAs} with updated look up tables that contain tabulated photon spectra including the radiative correction for a variety of WIMP masses (see appendix \ref{sec:micro}). These are formatted to replace the default spectra which only include an FSR component.

\section{Acknowledgements}
AS would like to thank Nordita and the OKC for the hospitality and resources that allowed a majority of the work to be undertaken.
The work of AS and AGW is supported by the Australian Research Council through the Centre of Excellence for Particle Physics at the Terascale CE110001004. MW is supported by the Australian Research Council Future Fellowship FT140100244. KF is supported by the Vetenskapsradet (Swedish Research Council).


\appendix
\section{Use of spectra in \micrOMEGAs}
\label{sec:micro}
The open source software \micrOMEGAs\: 4.1.8 contains lookup spectra tables for a variety of SM channels that it uses for its various array of calculations. These are located in the \texttt{/sources/data/} directory of the \micrOMEGAs\:  home directory as \texttt{.dat} files corresponding to a variety of SM decay channels. For example, the relevant tables for $\mu\mu\rightarrow X$, where $X$ are photons or electrons/positrons, are located in \texttt{mm.dat}. Each data file contains tabulated $E\frac{dN}{dE}$ in rows of $N_z=250$ elements corresponding to WIMP masses of {2, 5, 10, 25, 50, 80.3, 85, 91.2, 92, 95, 100, 110, 120, 125, 130, 140, 150,
176, 200, 250, 350, 500, 750, 1000, 1500, 2000, 3000 and 5000} GeV respectively. These default tables are used by \micrOMEGAs' internal routines to generate spectra for arbitrary user specified muon final states.   

The binning for these spectra is neither linear nor logarithmic. The i$^\text{th}$ energy bin for a spectrum corresponding to a DM mass $m_\text{DM}$ is given by 
$$
	E_i = m_\text{DM}\,e^{Z_i}\,,
$$
where 
$$
	Z_i = -7\ln(10)\left(\frac{i-1}{N_z}\right)^{1.5}\,. 
$$
The default \micrOMEGAs\: spectra are calculated using \tt{PYTHIA} 6.4\rm.
The new radiative correction discussed in section \ref{sec:spectra} was implemented into \micrOMEGAs\: by replacing the default decay spectra for the $\mu \mu \rightarrow \gamma$ with the correct binning. Since our radiative correction to the muon decay when generated with \tt{PYTHIA} \rm do not include spectra for WIMP masses $< 3$ GeV, the \micrOMEGAs\: source code needed to be changed accordingly; \micrOMEGAs '\: internal routines are told by default that the first spectrum in the lookup table \texttt{mm.dat} is for a WIMP mass of 2 GeV. This can be done by replacing the \texttt{/sources/spectra.c} file with the one provided in the arxiv version of this paper. We also include the modified look up table \texttt{mm\_radiative.dat} that includes the extra soft photons from the radiative correction down to a lower energy threshold of 0.1 MeV.

\bibliography{muons}
\bibliographystyle{JHEP}

\end{document}